\documentstyle[preprint,aps,epsfig]{revtex}
\begin{document}
\draft
\title{Computer simulations of electrorheological fluids \\
      in the dipole-induced dipole model}
\author{Y. L. Siu, Jones T. K. Wan and K. W. Yu}
\address{Department of Physics, The Chinese University of Hong Kong, \\
         Shatin, New Territories, Hong Kong, China}
\maketitle

\begin{abstract}
We have employed the multiple image method to compute the interparticle
force for a polydisperse electrorheological (ER) fluid in which the
suspended particles can have various sizes and different permittivites.
The point-dipole (PD) approximation being routinely adopted in computer 
simulation of ER fluids is shown to err considerably when the particles 
approach and finally touch due to multipolar interactions.
The PD approximation becomes even worse when the dielectric contrast
between the particles and the host medium is large.
From the results, we show that the dipole-induced-dipole (DID) model 
yields very good agreements with the multiple image results for a wide 
range of dielectric contrasts and polydispersity. 
As an illustration, we have employed the DID model to simulate the athermal 
aggregation of particles in ER fluids both in uniaxial and rotating fields. 
We find that the aggregation time is significantly reduced.
The DID model accounts for multipolar interaction partially and 
is simple to use in computer simulation of ER fluids.
\end{abstract}
\vskip 5mm

\section{Introduction}

The prediction of the yield stress for electrorheological (ER) fluids is
the main concern in theoretical investigations of ER fluids.
Early studies failed to derive the experimental yield stress data
\cite{Adriani} because these studies were almost based on a point-dipole
approximation \cite{Kling89,Kling91}.
The point-dipole approximation is routinely adopted in computer simulation
because it is simple and easy to use. Since many-body and multipolar
interactions between particles have been neglected, the strength of ER
effects predicted by this model is of an order lower than the experimental
results. Hence, substantial effort has been made to sort out more
accurate models.

Klingenberg and coworkers developed empirical force expression from
numerical solution of Laplace's equation \cite{Kling}.
Davis used the finite-element method \cite{Davis}.
Clercx and Bossis developed a full multipolar treatment to account for
multipolar polarizability of spheres up to 1,000 multipolar orders
\cite{Clercx}.
Yu and coworkers developed an integral equation method which avoids the
match of complicated boundary conditions on each interface of the particles
and is applicable to nonspherical particles and multimedia \cite{Yu}.
Although the above methods are accurate, they are relatively complicated
to use in dynamic simulation of ER fluids.
Alternative models have been developed to circumvent the problem:
the coupled-dipole model \cite{Tao} and the dipole-induced-dipole model
\cite{Yu2000}, which take care of mutual polarization effects when the 
particles approach and finally touch.

The DID model accounts for multipolar interactions partially and 
is simple to use in computer simulation of ER fluids \cite{Yu2000}.
As an illustration, we employed the DID model to simulate the athermal 
aggregation of particles in ER fluids both in uniaxial and rotating fields. 
We find that the aggregation time is significantly reduced.
In the next section, we review the multiple image method and establish the 
dipole-induced dipole (DID) model. In section III, we apply the DID model 
to the computer simulation of ER fluids in a uniaxial field. In section 
IV, we extend the simulation to athermal aggregation in rotating fields. 
Discussion on our results will be given. 

\section{Improved Multiple Image Method}

Here we briefly review the multiple images method \cite{Yu2000} and 
extend the method slightly to handle different dielectric constants.
Consider a pair of dielectric spheres, of radii $a$ and $b$, 
dielectric constants $\epsilon_1$ and $\epsilon_1'$ respectively, 
separated by a distance $r$. 
The spheres are embedded in a host medium of dielectric constant 
$\epsilon_2$. Upon the application of an electric field ${\bf E}_0$,
the induced dipole moment inside the spheres are respectively given by
(SI units):
\begin{equation}
p_{a0}=4\pi\epsilon_0 \epsilon_2 \beta E_0 a^3,\ \ \ \
  p_{b0}=4\pi\epsilon_0 \epsilon_2 \beta' E_0 b^3,
\end{equation}
where the dipolar factors $\beta,\beta'$ are given by:
\begin{equation}
\beta={\epsilon_1-\epsilon_2 \over \epsilon_1+2\epsilon_2},\ \ \
\beta'={\epsilon_1'-\epsilon_2 \over \epsilon_1'+2\epsilon_2}.
\end{equation}

From the multiple image method \cite{Yu2000}, 
the total dipole moment inside sphere $a$ is:
\begin{eqnarray}
p_{aT} &=& (\sinh \alpha)^3 \sum_{n=1}^\infty \left[
  {p_{a0} b^3 (-\beta)^{n-1}(-\beta')^{n-1}
    \over (b\sinh n\alpha + a\sinh (n-1)\alpha)^3}
    +{p_{b0} a^3 (-\beta)^{n}(-\beta')^{n-1} 
    \over (r \sinh n\alpha)^3 } \right],
\label{trans-a-dielectric} \\
p_{aL} &=& (\sinh \alpha)^3 \sum_{n=1}^\infty \left[
  {p_{a0} b^3 (2\beta)^{n-1}(2\beta')^{n-1}
  \over (b\sinh n\alpha + a\sinh (n-1)\alpha)^3}
    +{p_{b0} a^3 (2\beta)^{n}(2\beta')^{n-1} 
    \over (r \sinh n\alpha)^3 } \right],
\label{long-a-dielectric} 
\end{eqnarray}
where the subscripts $T$ ($L$) denote a transverse (longitudinal) field, 
i.e., the applied field is perpendicular (parallel) to the line joining 
the centers of the spheres. 
Similar expressions for the total dipole moment inside sphere $b$ can be
obtained by interchanging $a$ and $b$, as well as $\beta$ and $\beta'$. 
The parameter $\alpha$ satisfies:
$$
\cosh\alpha={r^2 - a^2 - b^2 \over 2 ab}.
$$
In Ref.\cite{Yu2000}, we checked the validity of these expressions by 
comparing with the integral equation method. 
We showed that these expression are valid at high contrast. 
Our improved expressions will be shown to be good at low contrast as well 
(see below).

The force between the spheres is given by \cite{Jackson}:
\begin{equation}
F_T = {E_0 \over 2}{\partial\over\partial r}(p_{aT}+p_{bT}),\ \ \ 
F_L = {E_0 \over 2}{\partial\over\partial r}(p_{aL}+p_{bL}).
\end{equation}

For monodisperse ER fluids ($a=b$, $\beta=\beta'$ and $p_a=p_b=p_0$), 
Klingenberg defined an empirical force expression \cite{Kling}:
\begin{eqnarray}
{{\bf F}\over F_{PD}} = (2K_\parallel \cos^2 \theta - K_\perp \sin^2 \theta)
  \hat{\bf r} + K_\Gamma \sin 2\theta \hat{\theta},
\end{eqnarray}
being normalized to the point-dipole force $F_{PD}=-3 p_{0}^2 /r^4$,
where $K_\parallel, K_\perp$ and $K_\Gamma$ (all tending to unity at large
separations) are three force functions being determined from the numerical
solution of Laplace's equation.
The Klingenberg's force functions can be shown to relate to our 
multiple image moments as follow (here $a=b$, $\beta=\beta'$ and $p_a=p_b$): 
\begin{eqnarray}
K_\parallel={1 \over 2}{\partial \tilde{p}_L \over \partial r}, \ \ \
K_\perp=-{\partial \tilde{p}_T \over \partial r}, \ \ \
K_\Gamma={1\over r}(\tilde{p}_T - \tilde{p}_L),
\label{Klingen}
\end{eqnarray}
where $\tilde{p}_L=p_L/F_{PD}E_0$ and $\tilde{p}_T=p_T/F_{PD}E_0$ are
the reduced multiple image moments of each sphere. 
We computed the numerical values of these force functions separately 
by the approximant of Table I of the second reference of 
Ref.\cite{Kling} and by Eq.(\ref{Klingen}). 

In Fig.1, we plot the multiple image results and the Klingenberg's 
empirical 
expressions. We show results for the perfectly conducting limit 
($\beta=1$) 
only. For convenience, we define the reduced separation $\sigma=r/(a+b)$.
For reduced separation $\sigma>1.1$, simple analytic expressions were 
adopted by Klingenberg. As evident from Fig.1, the agreement with the 
multiple image results is impressive at large reduced separation 
$\sigma>1.5$, for all three empirical force functions. 
However, significant deviations occur for $\sigma<1.5$, especially for 
$K_\parallel$.
For $\sigma\le 1.1$, alternative empirical expressions were adopted by 
Klingenberg. For $K_\perp$, the agreement is impressive, 
although there are deviations for the other two functions.
From the comparison, we would say that reasonable agreements have been 
obtained. Thus, we are confident that the multiple image expressions give
reliable results.

The analytic multiple image results can be used to compare among the 
various models according to how many terms are retained in the
multiple image expressions:
(a) point-dipole (PD) model: $n=1$ term only,
(b) dipole-induced-dipole (DID) model: $n=1$ to $n=2$ terms only, and
(c) multipole-induced-dipole (MID) model: $n=1$ to $n=\infty$ terms.

In a previous work \cite{Yu2000}, we examine the case of different size
but equal dielectric constant ($\beta=\beta'$) only.
Here we focus on the case $a=b$ and study the effect of different 
dielectric constants. 
In Fig.2, we plot the interparticle force in the longitudinal field case 
against the reduced separation $\sigma$ between the spheres for (a) 
$\beta=9/11$ ($\epsilon_1/\epsilon_2=10$) and 
(b) $\beta=1/3$ ($\epsilon_1/\epsilon_2=2$)
and various $\beta'/\beta$ ratios. At low contrast, the DID model almost 
coincides with the MID results.
In contrast, the PD model exhibits significant deviations.
It is evident that the DID model generally gives better results than PD for
all polydispersity.

\section{Computer simulation in the dipole-induced-dipole model}

The multiple image expressions [Eqs.(3)--(4)] allows us to calculate 
the correction factor defined as the ratio between the DID and PD forces: 
\begin{eqnarray}
{F_{DID}^{(\perp)} \over F_{PD}^{(\perp)}}
&=& 1 - {\beta a^3 r^5\over (r^2-b^2)^4} - {\beta' b^3 r^5\over (r^2-a^2)^4}
  + {\beta \beta' a^3 b^3 (3r^2-a^2-b^2)\over (r^2-a^2-b^2)^4}, \\
{F_{DID}^{(\parallel)} \over F_{PD}^{(\parallel)}}
&=& 1 + {2\beta a^3 r^5\over (r^2-b^2)^4} + {2\beta' b^3 r^5\over 
(r^2-a^2)^4}
  + {4\beta \beta' a^3 b^3 (3r^2-a^2-b^2)\over (r^2-a^2-b^2)^4}, \\
{F_{DID}^{(\Gamma)} \over F_{PD}^{(\Gamma)}}
&=& 1 + {\beta a^3 r^3\over 2(r^2-b^2)^3} + {\beta' b^3 r^3\over 
2(r^2-a^2)^3}
  + {3\beta \beta' a^3 b^3 \over (r^2-a^2-b^2)^3},
\end{eqnarray}
where $F_{PD}^{(\perp)}=3 p_{a0} p_{b0}/r^4$, 
$F_{PD}^{(\parallel)}=-6 p_{a0} p_{b0}/r^4$ and
$F_{PD}^{(\Gamma)}=-3 p_{a0} p_{b0}/r^4$ are the point-dipole forces for 
the transverse, longitudinal and $\Gamma$ cases respectively.
These correction factors can be readily calculated in computer simulation
of polydisperse ER fluids.
The results show that the DID force deviates significantly from the PD 
force at high contrast when $\beta$ and $\beta'$ approach unity. 
The dipole induced interaction will generally decrease (increase) the 
magnitude of the transverse (longitudinal) interparticle force with 
respect to the point-dipole limit.

For simplicity, we consider the case of two equal spheres of radius $a$, 
initially at rest and at a separation $d_0$.
An electric field is applied along the line joining the centers of the 
sphere. The equation of motion is given by:
\begin{equation}
{dz\over dt} = F_\parallel (2z),
\end{equation}
where $z$ is the displacement of one sphere from the center of mass.
The separation between the two spheres is thus $d=2z$ and the initial 
condition is $d=d_0$ at $t=0$.
Eq.(11) is a dimensionless equation. We have chosen the following natural 
scales to define the dimensionless variables:
$$
{\rm length}\ \sim a,\ \ 
{\rm time}\ = t_0 \sim \frac{6 \pi \eta_{c} a^{2}}{F_{0}},\ \ 
{\rm force}\ = F_0 \sim \frac{\epsilon_{0} 
\epsilon_{2}^{2} a^{2} E_{0}^{2}}{4 \pi},
$$
where $E_0$ is the field strength, $m$ is the masss, $\eta_c$ is the 
coefficient of viscosity.
Using typical parameters, we find $t_0$ is of the order milliseconds.
We have followed Klingenberg \cite{Kling89,Kling91} to ignore the 
inertial effect, captured by the parameter $G$.
$$
G = \frac{\epsilon_{0} m \epsilon_{2}^{2} E_{0}^{2}}{144 \pi^{3} 
\eta_{c}^{2} a},
$$
The neglect of $G$ can be justified as follows.
For values common to ER suspension: $\eta_{c} \approx 0.1$ Pa s, 
$m \approx 8 \times 10^{-13}$ kg, $a \approx 5 \times 10^{-6}$m 
$\cite{Kling89}$, The inertial term $G$ is of the order $10^{-8}$.
We also neglect the thermal motion of the particles which is a valid 
assumption at high fields.
We should remark that the initial separation $d_0$ is related to the volume 
fraction $\phi$, defined as the ratio of the volume of the sphere to 
that of the cube which contains the sphere \cite{Kling91}, i.e.
$\phi = V_{\rm sphere}/V_{\rm cube}$, and
$$
\frac{d_{0}}{2a} = \left( {\frac{\pi}{6 \phi}} \right)^{1/3}.
$$
For the PD approximation, Eq.(11) admits an analytic solution: 
\begin{equation}
z=\left[ \left({d_0\over 2}\right)^5 - {15p_0^2t\over 8} \right]^{1/5}.
\end{equation}
We integrate the equation of motion by the 4th order Runge-Kutta 
algorithm, with time steps $\delta t=0.01$ and 0.001 for small and
large volume fractions respectively. 
In Fig.3(a), we plot the displacement $d/a$ versus time graph for the PD 
case and find excellent agreement between analytic and numerical results.

For the DID model, we have to integrate the equation of motion numerically.
In Fig.4, we plot the displacement $d/a$ versus time graph for the 
aggregation of two spheres in uniaxial fields. At small volume fractions, 
i.e., when the initial separation is large, the time for aggregation is 
large and the DID results deviate slightly from the PD results.
However, at large volume fractions, the DID results are significantly smaller
than the PD calculations. The effect becomes even more pronounced at large
$\beta$.

In Fig.5(a), we plot the ratio of aggregation time of the DID to PD cases.
The results showed clearly that the aggregation time has been 
significantly reduced when mutual polarization effects are considered. 
The reduction in aggregation time becomes even pronounced for small 
initial separations. 

\section{Athermal aggregation in the rotating field}

Recently, Martin and coworkers \cite{Martin} demonstrated athermal
aggregation with the rotating field. When a rotating field is applied in the 
$x$-$y$ plane at a sufficiently high frequency that particles do not move 
much in one period, an average attractive dipolar interaction is created. 
The result of this is the formation of plates in the $x$-$y$ plane.
Consider a rotating field applied in the $x$-$y$ plane:
$E_x=E_0 \cos \omega t, E_y=E_0 \sin \omega t$.
The dimensionless equation of motion for the two sphere case becomes: 
\begin{equation}
{dx\over dt} = F_\parallel \cos^2 \omega t + F_\perp \sin^2 \omega t,\ \ 
{dy\over dt} = -F_\Gamma \sin 2\omega t, 
\end{equation}
where $(x,y)$ is the displacement of one sphere from the center of mass.
For large $\omega$, we may safely neglect the $y$ component of the motion.
In the PD approximation, $F_\parallel=-6p_0^2/r^4$ and 
$F_\perp=3p_0^2/r^4$, we find the analytic result: 
\begin{equation}
x=\left[ \left({d_0\over 2}\right)^5 - {15p_0^2\over 64\omega}
  (2\omega t + 3\sin 2\omega t) \right]^{1/5}.
\end{equation}
The separation between the two spheres is just $d=2x$, with the initial 
separation $d=d_0$ at $t=0$. 
In the rotating field case, we also integrate the equation of motion by 
the 4th order Runge-Kutta algorithm, but with $\delta t = 1/(4\omega)$ 
and $1/(40\omega)$ as the time steps. Note that $\delta t=1/(4\omega)$ 
should be the largest time step which can be used because we must at least 
go through a cycle consisting of the transverse and longitudinal field 
cases. The oscillating effect of a rotating field is less observable 
when the time step is smaller than this maximum value.

In Fig.3(b), we plot the displacement versus time graph for the PD 
case in rotating field and find an excellent agreement between analytic 
and numerical results. It is evident that the aggregation time is 4 times 
of that of the uniaxial field case. In fact, Eq.(14) reduces to Eq.(12) 
as $\omega \to 0$. However, at large $\omega$, Eq.(14) becomes: 
$$
x=\left[ \left({d_0\over 2}\right)^5 - {15p_0^2t\over 32} \right]^{1/5}.
$$
That is, in the PD approximation, the time average force becomes 1/4 of 
that of the uniaxial field case. It is because the two dipole moments 
spend equal times in the transverse and longitudinal orientations, while 
$F_\parallel = -2 F_\perp$ in the PD case, leading to an overall 
attractive force that is 1/4 of the force of the uniaxial field case. 
When the multiple image force is included, we expect that the magnitude of 
$F_\parallel$ increases while that of $F_\perp$ decreases and we expect
an even larger attractive force when the spheres approach. In this case, 
the aggregation time must be reduced even more significantly.

In Fig.5(b), we plot the ratio of aggregation time of the DID to PD cases
for $\beta=1$ and several $\omega$. The $\omega=0$ curve is just for the 
uniaxial field case. 
The results showed clearly that the aggregation time has been 
significantly reduced when mutual polarization effects are considered. 
The reduction in aggregation time becomes even pronounced for small 
initial separations. 
It is observed that fluctuations exist when the initial separation
between the spheres is 2.4$a$ or less. 
It is because the motion is sensitive to the initial orientation of the 
dipoles when the spheres are too close.

Similarly, we consider the aggregations of 3 and 4 equal spheres,
arranged in a chain, an equilateral triangle and a square. 
For a chain of 3 spheres in a rotating field, the central sphere does not 
move, while the two spheres at both ends move towards the central sphere. 
For 3 spheres in an equilateral triangle, the center of mass (CM) will 
not move while each sphere moves towards the CM, subject to the force of the 
other two spheres. The same situation occurs for 4 spheres in a square, 
in which each sphere moves towards the CM, subject to the force of the other 
3 spheres.

In the PD approximation, we report the analytic results as follows.
For 3 spheres in a chain,
\begin{equation}
x = \left[d_0^5-\frac{255p^{2}_{0}}{64\omega}
(2\omega t+3\sin 2\omega t)\right]^{1/5}.
\end{equation}
For 3 spheres in an equilateral triangle,
\begin{equation}
x = \left[\left(\frac{d_0}{\sqrt{3}}\right)^{5}-\frac{5p^{2}_{0}}
{8\sqrt{3}\omega}
(4\omega t-\sin 2\omega t)\right]^{1/5}.
\end{equation}
For 4 spheres in a square,
\begin{equation}
x = \left[\left(\frac{d_0}{\sqrt{2}}\right)^{5}-\frac{15p^{2}_{0}}
{8\omega}
\left(\frac{4\sqrt{2}+1}{4}\omega t-\frac{8\sqrt{2}-3}{8}\sin 2\omega t
\right) \right]^{1/5}.
\end{equation}
In each of the above cases, $x$ is the distance of one sphere from the
center of mass. In the case of 3 spheres in a chain, the separation
between the spheres is same as $d=x$.
In the case of 3 spheres in an equilateral triangle, the separation 
between spheres is $d=\sqrt{3}x$.
In the case of 4 spheres in a square, the separation between spheres is
$d=\sqrt{2}x$.
Again, we integrate the equation of motion by the 4th order Runge-Kutta 
algorithm. We find excellent agreements between the analytic and numerical 
results (not shown here).

It has been found that the displacement in the $y$-direction is about 0.5\% 
for $\omega=5$ and a larger $\omega$ has been used in the simulation.
On the other hand, it is time consuming for simulations with $\omega>20$. 
It is evident from the displacement-time graph that the results are correct.
In Fig.6, the oscillation amplitude is reduced when the rotating frequency 
increases in the simulation. This is consistent with the assumption made in
our analytic expressions. 
In Fig.7, it is observed that fluctuations exist when the initial separation
between the spheres is 2.4$a$ or less in all three graphs. 
Again, it is because the motion is sensitive to the orientation of the 
dipoles when the spheres are close.
From the simulation, the reduction effects become even pronounced for the 
rotating electric field case than the uniaxial field case. 

\section*{Discussion and Conclusion}

Here a few comments on our results are in order.
In this work, we studied the aggregation time for several particles. 
We should also examine the morphology of aggregation, due to multiple 
image forces. In this connection, we can also examine the structural 
transformation by applying the uniaxial and rotating fields 
simultaneously \cite{Lo}.

We have done simulation in the monodisperse case.
Real ER fluids must be polydisperse in nature: the suspending particles 
can have various sizes or different permittivities.
Polydisperse electrorheological (ER) fluids have attracted considerable
interest recently because the size distribution and dielectric properties
of the suspending particles can have significant impact on the ER response
\cite{Ota}. We should extend the simulation to polydisperse case by using 
the DID model.

In summary, we have used the multiple image to compute the interparticle 
force for a polydisperse electrorheological fluid. 
We apply the formalism to a pair of spheres of different dielectric 
constants and calculate the force as a function of the separation.
The results show that the point-dipole approximation is oversimplified. 
It errs considerably because many-body and multipolar interactions are 
ignored.
The dipole-induced-dipole model accounts for multipolar interactions 
partially and yields overall satisfactory results in computer simulation 
of ER fluids while it is easy to use.

\section*{Acknowledgments}

This work was supported by the Research Grants Council of the Hong Kong 
SAR Government under grant CUHK4284/00P.

\begin{figure}[h]
\caption{Comparison of multiple image results with Klingenberg's 
 force functions.}
\end{figure}

\begin{figure}[h]
\caption{Force for longitudinal fielda: (a) $\beta$=9/11, (b) $\beta$=1/3.}
\end{figure}

\begin{figure}[h]
\caption{Comparison between analytic and simulation results in athermal 
 aggregation: (a) uniaxial field, (b) rotating field.}
\end{figure}

\begin{figure}[h]
\caption{Diplacement-time graph for athermal aggregation of two spheres
 in uniaxial field.}
\end{figure}

\begin{figure}[h]
\caption{Reduction of aggregation time for: (a) uniaxial field,
 (b) rotating field.}
\end{figure}

\begin{figure}[h]
\caption{Diplacement-time graph for athermal aggregation of two spheres
 in rotating field.}
\end{figure}

\begin{figure}[h]
\caption{Reduction of aggregation time for three and four spheres 
 in rotating field.}
\end{figure}

\newpage
\centerline{\epsfig{file=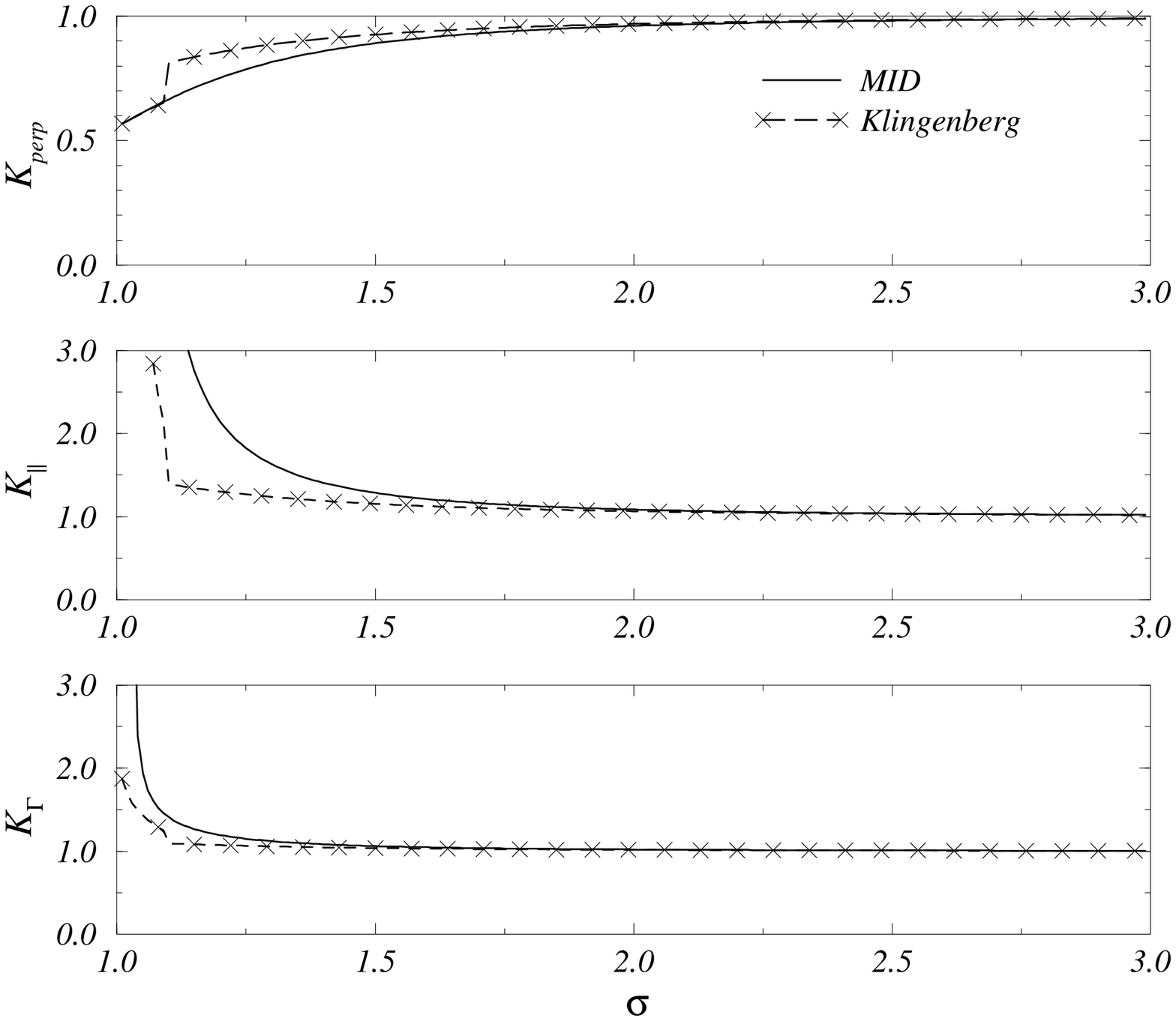,width=\linewidth}}
\centerline{Fig.1}

\newpage
\centerline{\epsfig{file=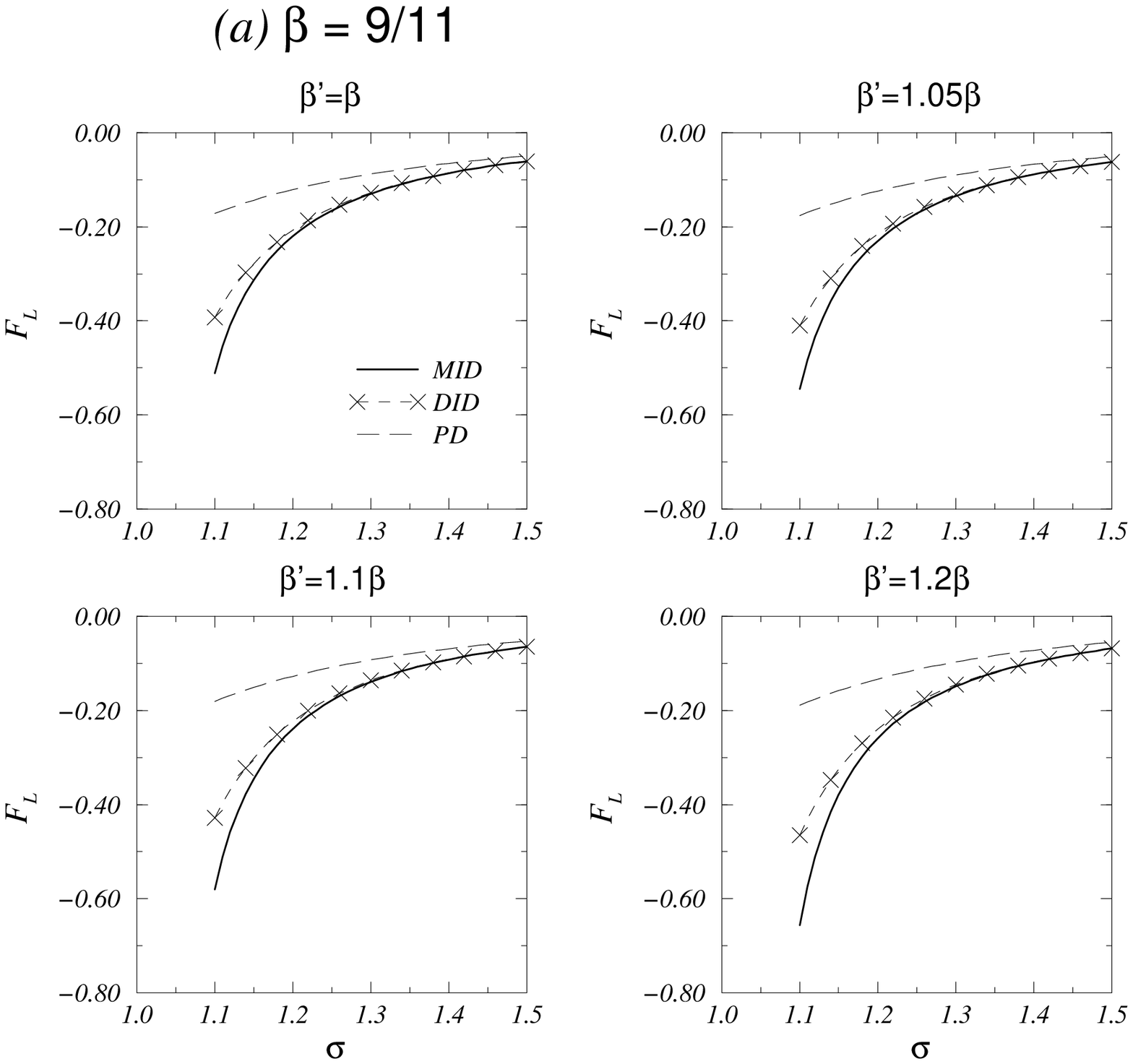,width=\linewidth}}
\centerline{Fig.2(a)}
\centerline{\epsfig{file=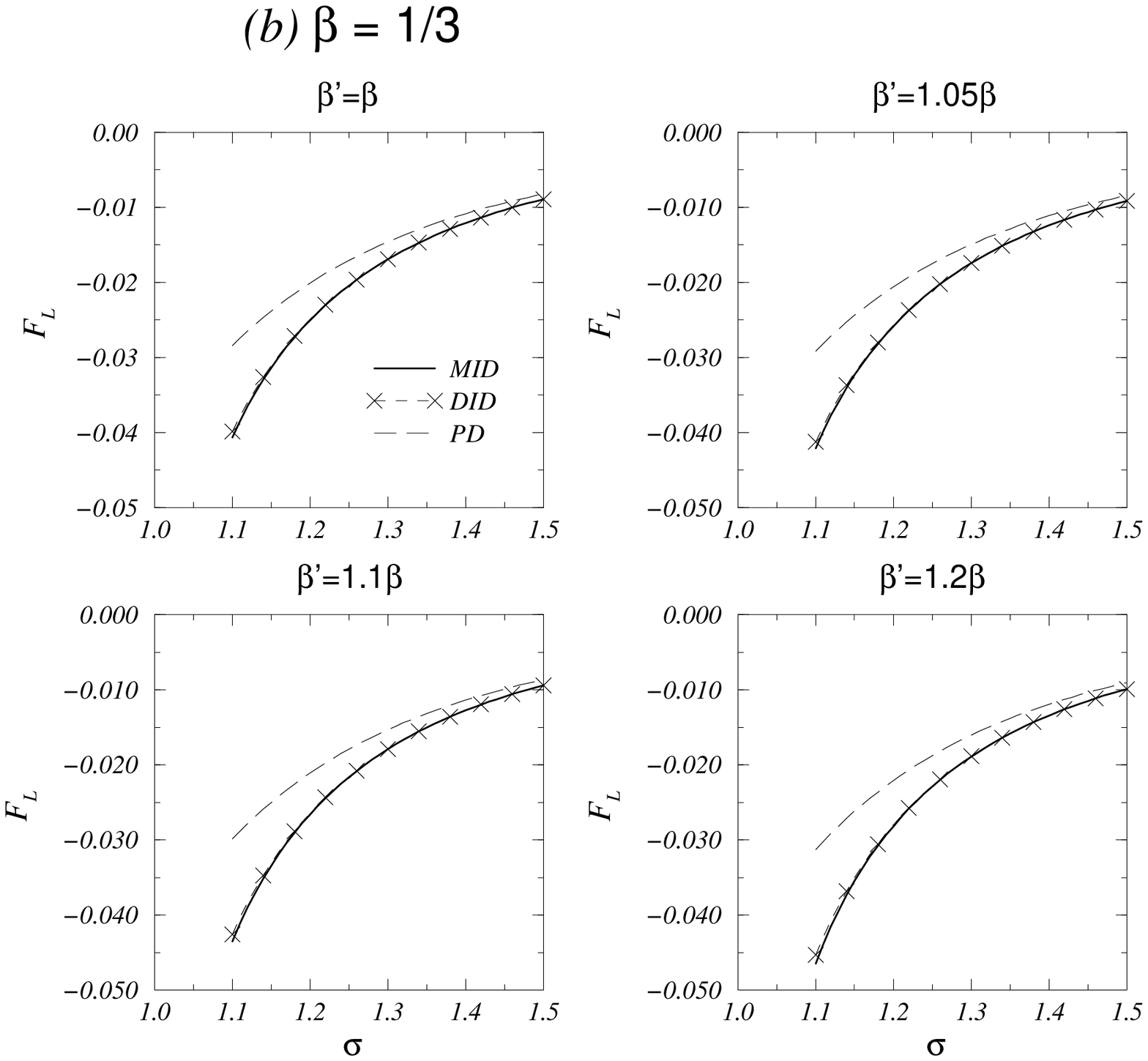,width=\linewidth}}
\centerline{Fig.2(b)}

\newpage
\centerline{\epsfig{file=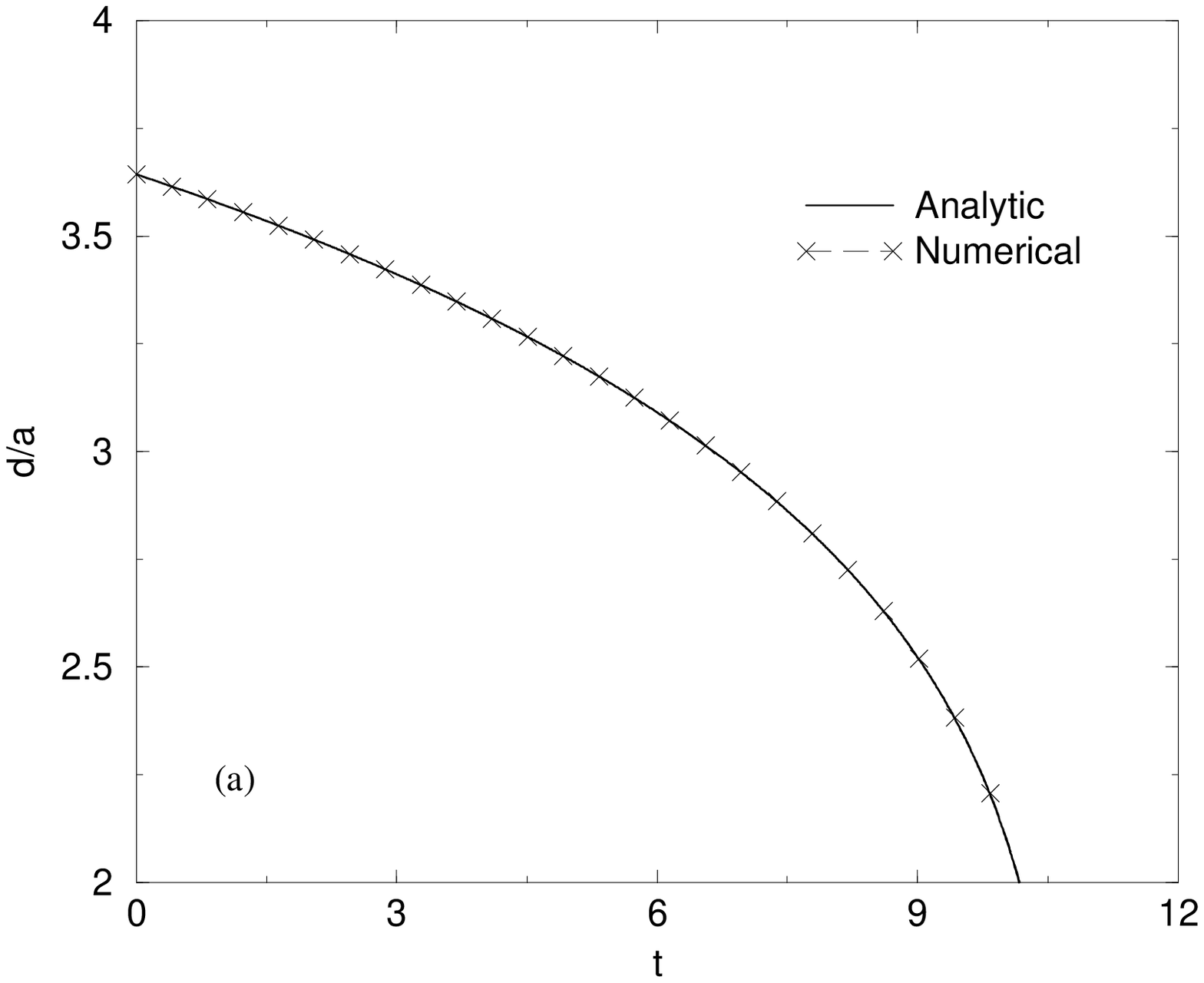,width=\linewidth}}
\centerline{Fig.3(a)}
\centerline{\epsfig{file=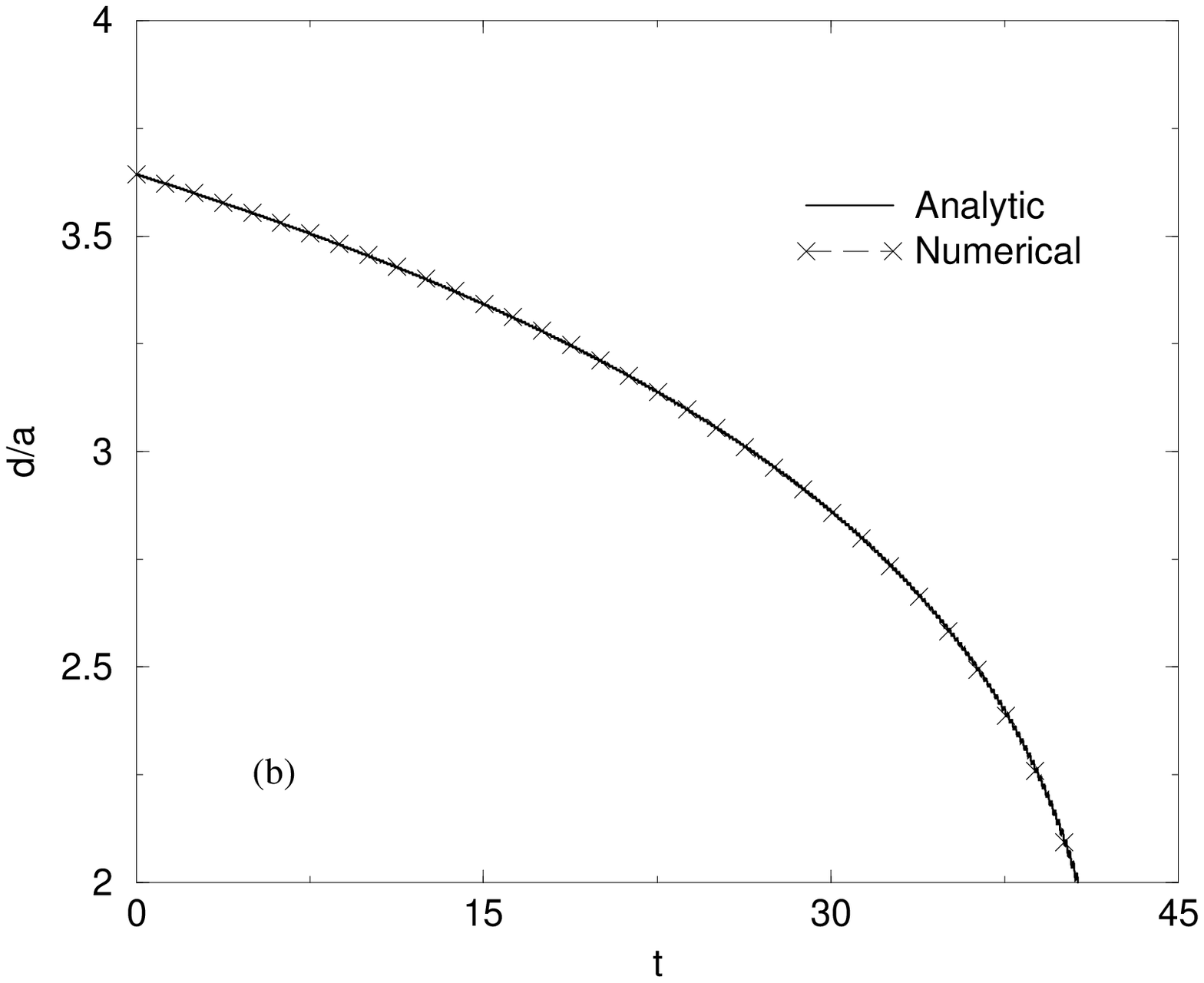,width=\linewidth}}
\centerline{Fig.3(b)}

\newpage
\centerline{\epsfig{file=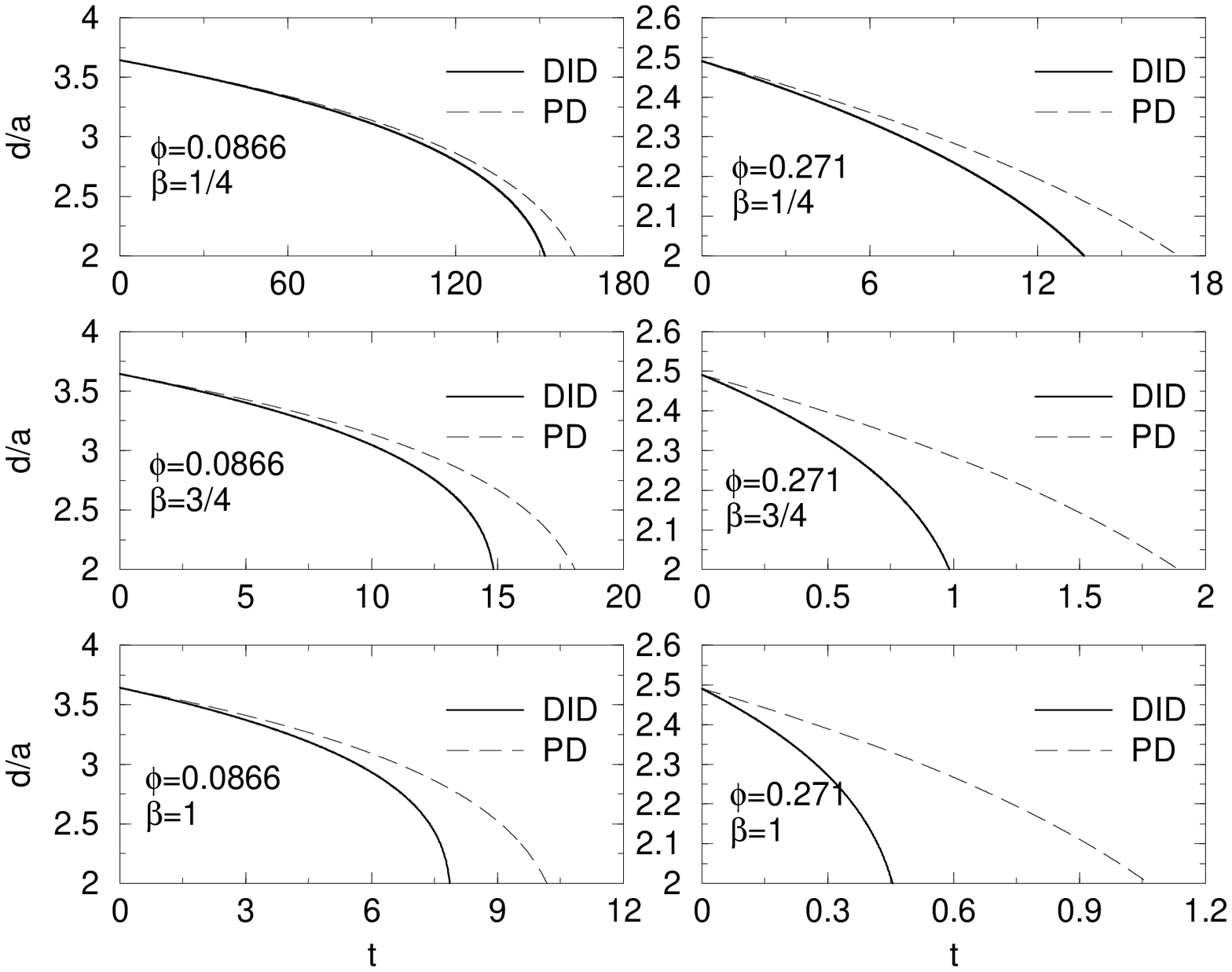,width=\linewidth}}
\centerline{Fig.4}

\newpage
\centerline{\epsfig{file=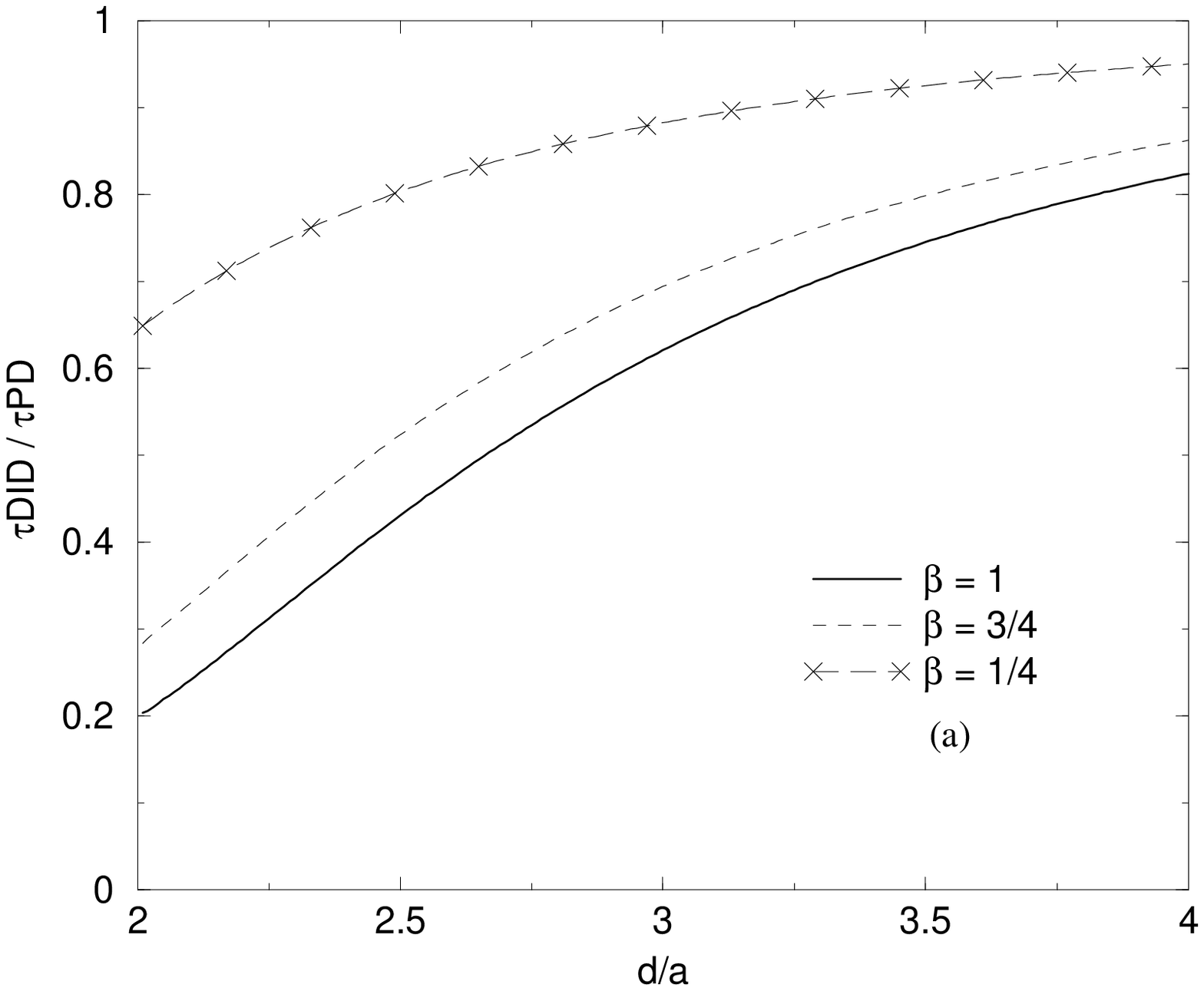,width=\linewidth}}
\centerline{Fig.5(a)}
\centerline{\epsfig{file=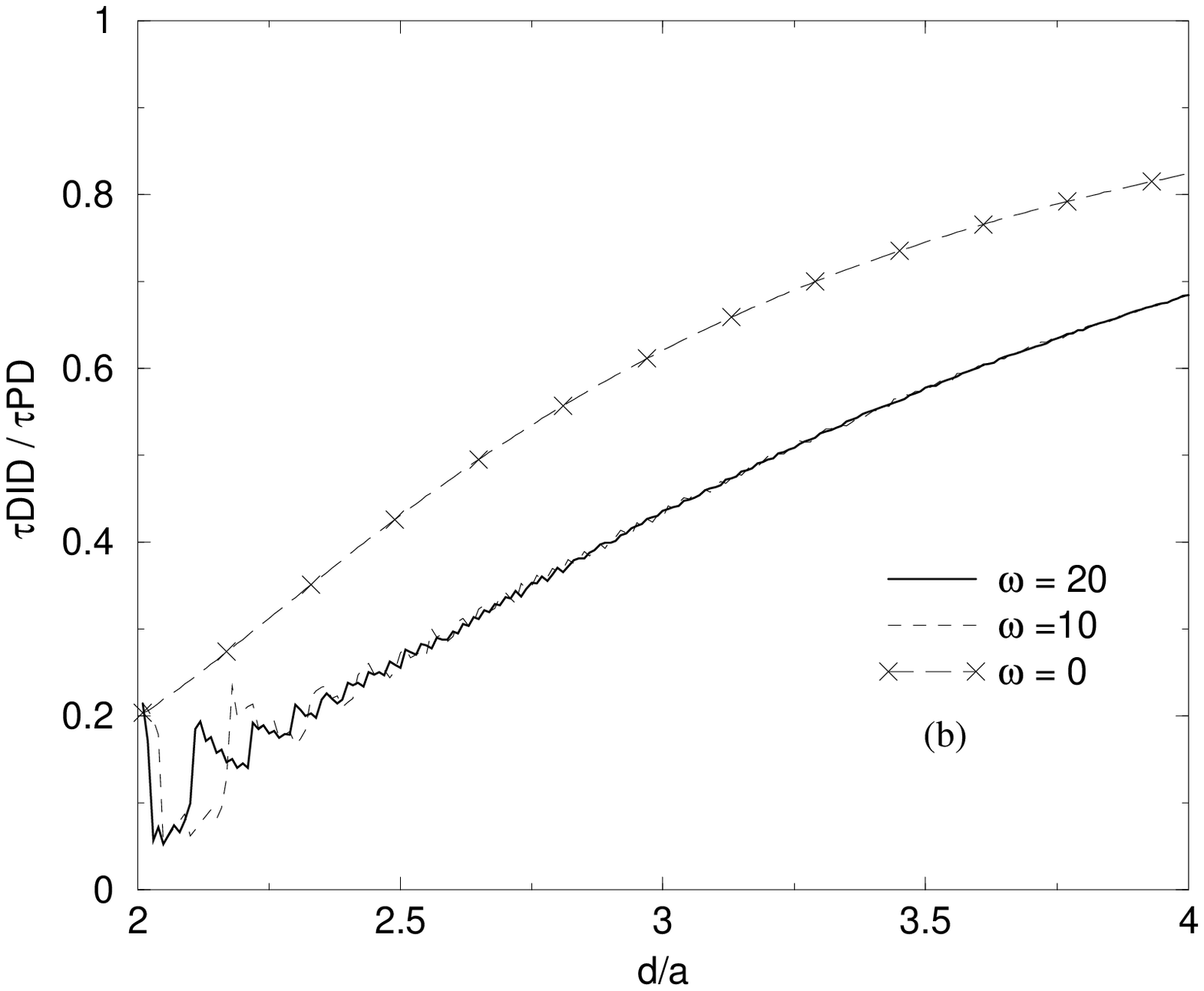,width=\linewidth}}
\centerline{Fig.5(b)}

\newpage
\centerline{\epsfig{file=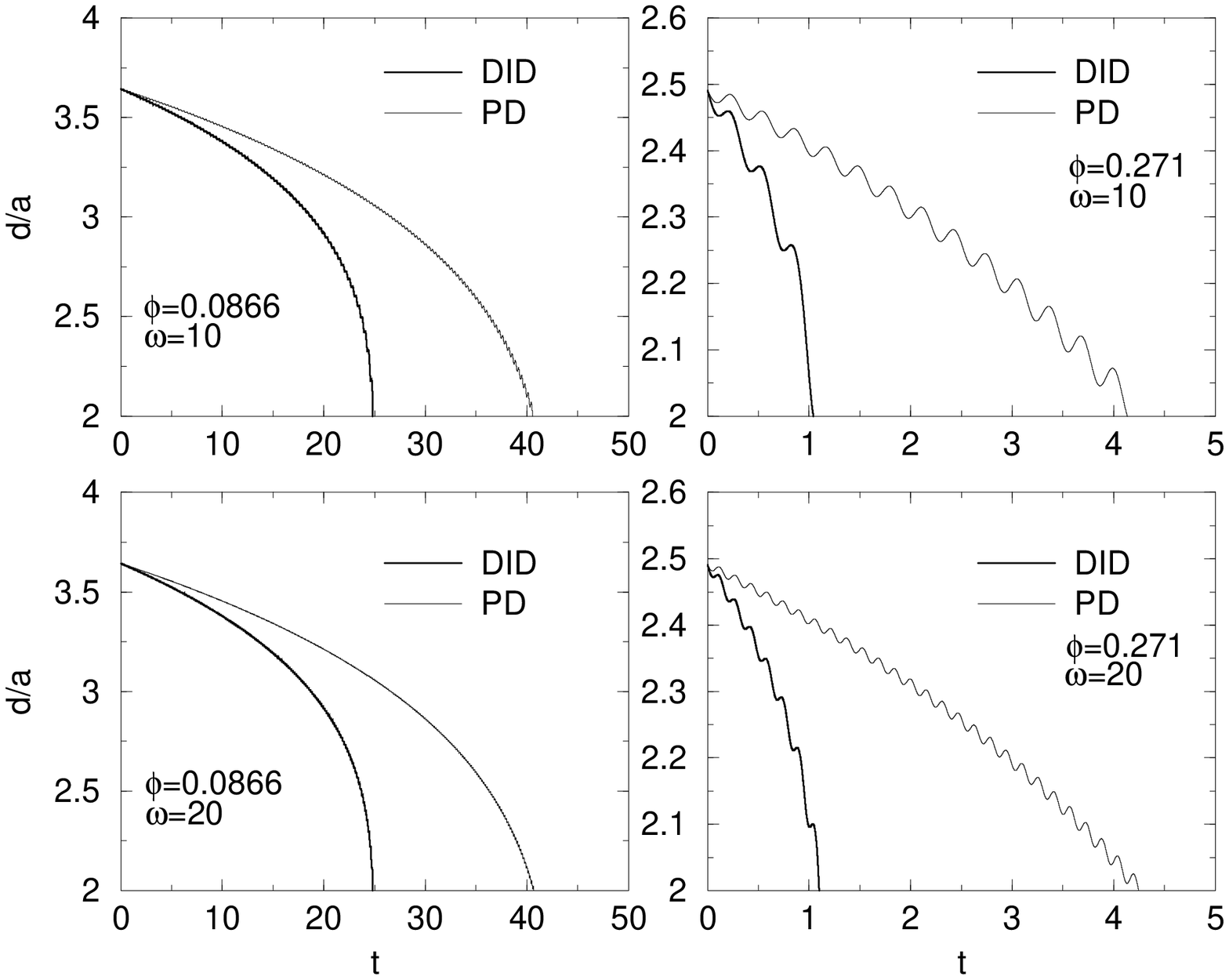,width=\linewidth}}
\centerline{Fig.6}

\newpage
\centerline{\epsfig{file=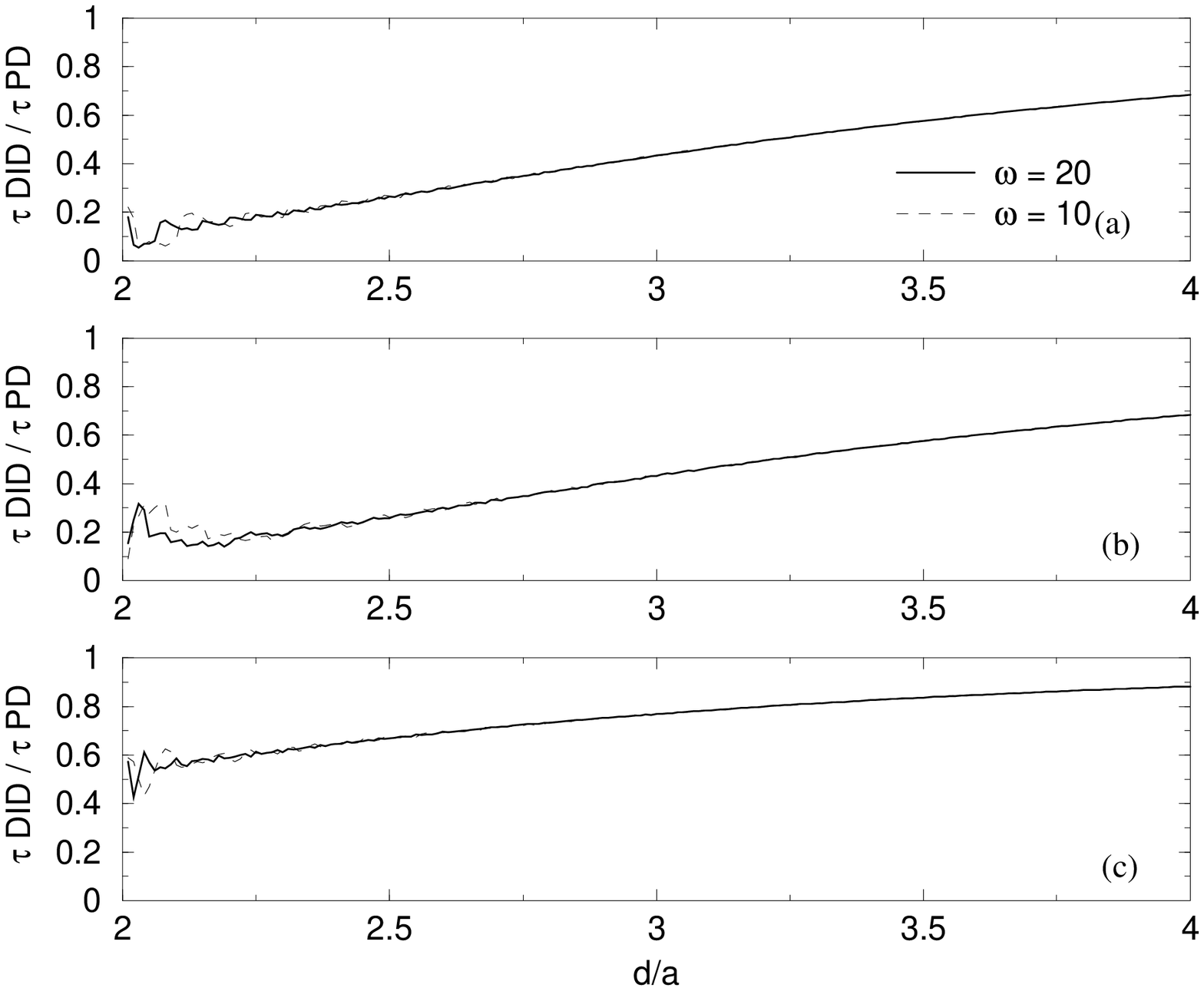,width=\linewidth}}
\centerline{Fig.7}

\end{document}